\documentclass[prl,aps,twocolumn,notitlepage,superscriptaddress,showpacs,nofootinbib]{revtex4-1}

\usepackage{enumerate,appendix}
\usepackage{amsmath, amsthm, amssymb,commath,braket}
\usepackage{color,calc,graphicx}
\usepackage[usenames,dvipsnames,svgnames,table,cmyk,hyperref]{xcolor}
\usepackage[colorlinks]{hyperref}
\hypersetup{
	colorlinks = true,
	citecolor = {blue},
	linkcolor = {blue}
}

\usepackage[charter,cal=cmcal,sfscaled=false]{mathdesign}
\usepackage{booktabs}
\usepackage{multirow}
\usepackage{subfigure}
\usepackage{dcolumn}
\usepackage{mathrsfs}

 \newcommand{\trace}{{\rm Tr}}
 
\makeatletter

\newcommand{\Rmnum}[1]{\expandafter\@slowromancap\romannumeral #1@}
\makeatother

\begin{document}
\title{Photonic realization of quantum resetting}


\author{Zheng-Da Li}
\altaffiliation{These authors contributed equally to this work.}
\affiliation{Shanghai Branch, National Laboratory for Physical Sciences at Microscale and Department of Modern Physics, University of Science and Technology of China, Shanghai 201315, China}
\affiliation{CAS Center for Excellence and Synergetic Innovation Center in Quantum Information and Quantum Physics, University of Science and Technology of China, Shanghai 201315, China}

\author{Xu-Fei Yin}
\altaffiliation{These authors contributed equally to this work.}
\affiliation{Shanghai Branch, National Laboratory for Physical Sciences at Microscale and Department of Modern Physics, University of Science and Technology of China, Shanghai 201315, China}
\affiliation{CAS Center for Excellence and Synergetic Innovation Center in Quantum Information and Quantum Physics, University of Science and Technology of China, Shanghai 201315, China}

\author{Zizhu Wang}
\affiliation{Institute of Fundamental and Frontier Sciences, University of Electronic Science and Technology of China, Chengdu 610054, China}

\author{Li-Zheng Liu}
\affiliation{Shanghai Branch, National Laboratory for Physical Sciences at Microscale and Department of Modern Physics, University of Science and Technology of China, Shanghai 201315, China}
\affiliation{CAS Center for Excellence and Synergetic Innovation Center in Quantum Information and Quantum Physics, University of Science and Technology of China, Shanghai 201315, China}

\author{Rui Zhang}
\affiliation{Shanghai Branch, National Laboratory for Physical Sciences at Microscale and Department of Modern Physics, University of Science and Technology of China, Shanghai 201315, China}
\affiliation{CAS Center for Excellence and Synergetic Innovation Center in Quantum Information and Quantum Physics, University of Science and Technology of China, Shanghai 201315, China}

\author{Yu-Zhe Zhang}
\affiliation{Shanghai Branch, National Laboratory for Physical Sciences at Microscale and Department of Modern Physics, University of Science and Technology of China, Shanghai 201315, China}
\affiliation{CAS Center for Excellence and Synergetic Innovation Center in Quantum Information and Quantum Physics, University of Science and Technology of China, Shanghai 201315, China}

\author{Xiao Jiang}
\affiliation{Shanghai Branch, National Laboratory for Physical Sciences at Microscale and Department of Modern Physics, University of Science and Technology of China, Shanghai 201315, China}
\affiliation{CAS Center for Excellence and Synergetic Innovation Center in Quantum Information and Quantum Physics, University of Science and Technology of China, Shanghai 201315, China}

\author{Jun Zhang}
\affiliation{Shanghai Branch, National Laboratory for Physical Sciences at Microscale and Department of Modern Physics, University of Science and Technology of China, Shanghai 201315, China}
\affiliation{CAS Center for Excellence and Synergetic Innovation Center in Quantum Information and Quantum Physics, University of Science and Technology of China, Shanghai 201315, China}

\author{Li Li}
\affiliation{Shanghai Branch, National Laboratory for Physical Sciences at Microscale and Department of Modern Physics, University of Science and Technology of China, Shanghai 201315, China}
\affiliation{CAS Center for Excellence and Synergetic Innovation Center in Quantum Information and Quantum Physics, University of Science and Technology of China, Shanghai 201315, China}

\author{Nai-Le Liu}
\affiliation{Shanghai Branch, National Laboratory for Physical Sciences at Microscale and Department of Modern Physics, University of Science and Technology of China, Shanghai 201315, China}
\affiliation{CAS Center for Excellence and Synergetic Innovation Center in Quantum Information and Quantum Physics, University of Science and Technology of China, Shanghai 201315, China}

\author{Xiao-Bo Zhu}
\affiliation{Shanghai Branch, National Laboratory for Physical Sciences at Microscale and Department of Modern Physics, University of Science and Technology of China, Shanghai 201315, China}
\affiliation{CAS Center for Excellence and Synergetic Innovation Center in Quantum Information and Quantum Physics, University of Science and Technology of China, Shanghai 201315, China}

\author{Feihu Xu}
\affiliation{Shanghai Branch, National Laboratory for Physical Sciences at Microscale and Department of Modern Physics, University of Science and Technology of China, Shanghai 201315, China}
\affiliation{CAS Center for Excellence and Synergetic Innovation Center in Quantum Information and Quantum Physics, University of Science and Technology of China, Shanghai 201315, China}

\author{Yu-Ao Chen}
\affiliation{Shanghai Branch, National Laboratory for Physical Sciences at Microscale and Department of Modern Physics, University of Science and Technology of China, Shanghai 201315, China}
\affiliation{CAS Center for Excellence and Synergetic Innovation Center in Quantum Information and Quantum Physics, University of Science and Technology of China, Shanghai 201315, China}

\author{Jian-Wei Pan}
\affiliation{Shanghai Branch, National Laboratory for Physical Sciences at Microscale and Department of Modern Physics, University of Science and Technology of China, Shanghai 201315, China}
\affiliation{CAS Center for Excellence and Synergetic Innovation Center in Quantum Information and Quantum Physics, University of Science and Technology of China, Shanghai 201315, China}


\begin{abstract}
Contrary to the usual assumption of at least partial control of quantum dynamics, a surprising recent result proved that an arbitrary quantum state can be probabilistically reset to a state in the past by having it interact with probing systems in a consistent, but \emph{uncontrolled} way. We present a photonic implementation to achieve this resetting process, experimentally verifying that a state can be probabilistically reset to its past with a fidelity of $0.870\pm0.012$. We further demonstrate the preservation of an entangled state, which still violates a Bell inequality, after half of the entangled pair was reset. The ability to reset uncontrolled quantum states has implications in the foundations of quantum physics and applications in areas of quantum technology.
\end{abstract}

\maketitle

In quantum theory, the words ``dynamics'' and ``control'' often go hand in hand. Indeed, the field of quantum control~\cite{RabitzQuantumControl} aims to optimize certain objectives by using an external control field to fine-tune the dynamics. On the other hand, altering the dynamics of a quantum system while having little or no control over it also has great theoretical and experimental implications. The \emph{refocusing} techniques, from spin echo~\cite{PhysRev.80.580,freeman1998spin} to dynamical decoupling~\cite{PhysRevLett.82.2417,zanardi}, use fast pulses to average out the effect of the environment, thus ``freezing'' the state of the target quantum system. Another related technique, universal refocusing~\cite{1602.07963}, goes even further and allows the system to go back in time. These techniques, however, assume at least some control over the dynamics. Theoretically, it is also possible to time-translate a quantum system, allowing the dynamics to go both forwards and backwards in time, with only minimal assumptions on the Hamiltonian~\cite{PhysRevLett.64.2965}. However, it is unlikely such a protocol can be realized, because of its astronomically small probability of success.

Using a theoretical tool called central matrix polynomials~\cite{drensky_formanek_book}, a recent family of protocols combines the strength of the techniques above to allow the \emph{uncontrolled} dynamics of a target quantum system to be neutralized, \emph{resetting} it to an earlier state~\cite{NavascuesPRX2018}. By ``uncontrolled" we mean that both the target system's free evolution and its joint interaction with other systems are unknown. By ``resetting'' we mean that the effects of the uncontrolled dynamics on the target system are canceled without {explicitly implementing the reverse evolution (cf. a recent protocol which reverses the dynamics by explicitly constructing the exact inverse unitary~\cite{PhysRevLett.123.210502})}, and an arbitrary state is reset to a state in the past. In these quantum resetting protocols~\cite{NavascuesPRX2018}, the target is an arbitrary quantum state $\mathcal{S}$ of dimension $d_{\mathcal{S}}$. It evolves freely according to a time-independent unknown Hamiltonian $\mathcal{H}_0$ for a time $t=T$. The goal is to probabilistically reset the target to its state at $t=0$ by making it interact with several probes, with the assumption that the target-probe interaction $V_\mathcal{SP}$ is always the same but otherwise unknown. In general, the number of probes needed scales as $O(d_{\mathcal{S}}^3)$~\cite{NavascuesPRX2018}. When the target is a qubit ($d_{\mathcal{S}}=2$), as in the experiment, the number of probes needed is at least four. 

\begin{figure*}
	\centering
	\includegraphics[width=\linewidth]{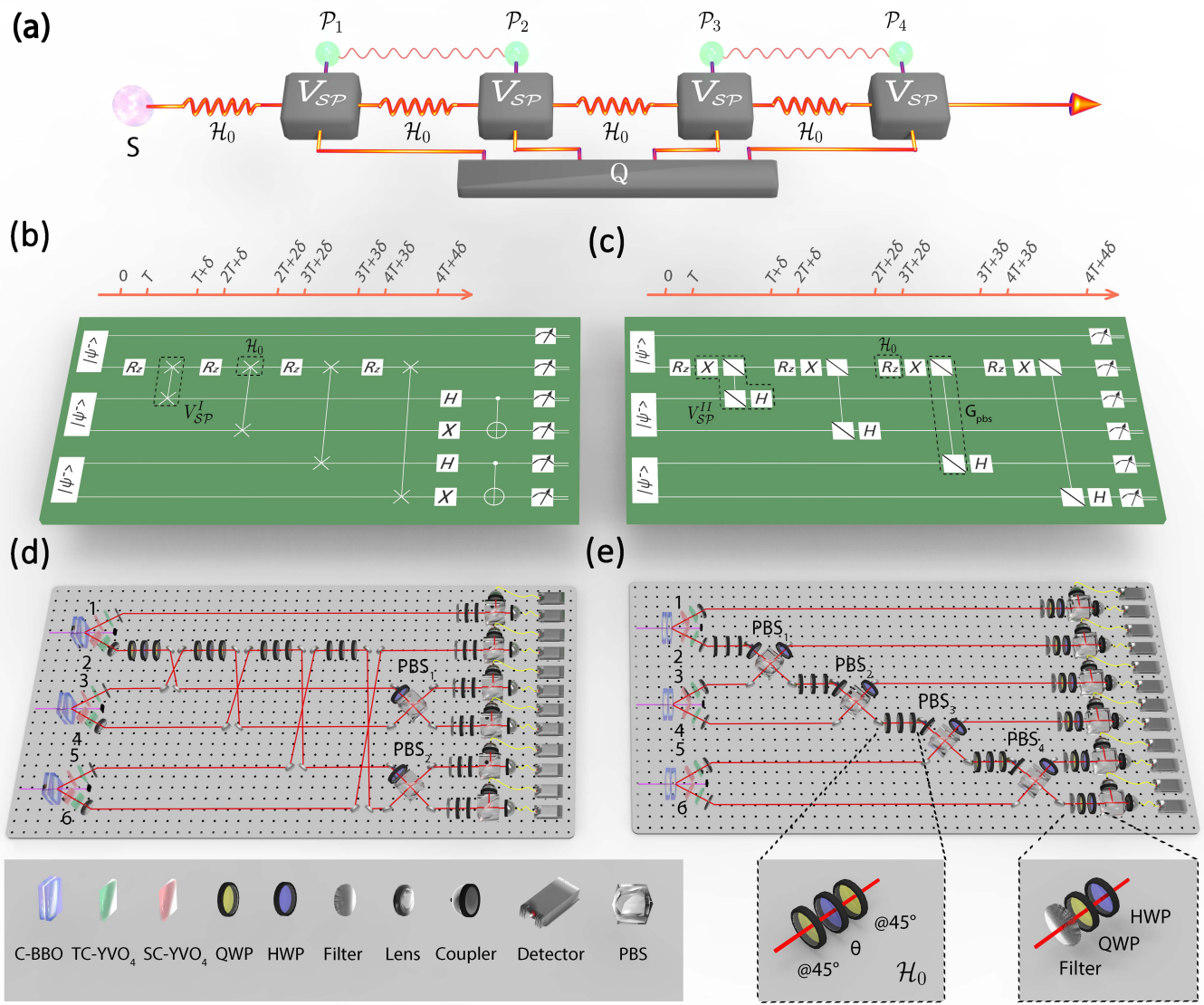}
	\caption{\textbf{Illustration of the protocol and corresponding experimental setup.} (a) Schematic diagram of the quantum resetting protocol for a single photon. A target photon (qubit) $S$, initialized at an arbitrary state, evolves over four unknown time-independent Hamiltonians $\mathcal{H}_0$ and interacts with four probes. A measurement result in Q heralds target's reset to its initial state. (b) Quantum circuit I based on SWAP gates. (c) Quantum circuit II based on Hadamard and non-unitary PBS gates. (d,e) Photonic implementation for circuit I and circuit II. A pulsed ultraviolet laser successively pumps {three sandwich-like combinations of $\beta$-barium borate crystals (C-BBO)} to generate three entangled photon pairs. Photon 2, heralded by photon 1, serves as the target system, while photons 3-6 serve as the probes. These photons pass through the free evolutions and the interactions, and finally they are detected by projection measurements. {SC-YVO$_4$ and TC-YVO$_4$ represent for spatial compensation (SC) and temporal compensation (TC) Yttrium Orthovanadate (YVO$_4$) crystals. QWP and HWP represent for quarter- and half-wave plates, respectively.}
	}
	\label{fig:setup}
\end{figure*}

Using linear-optical circuits and entangled photon pairs, we successfully neutralized the uncontrolled dynamics of a target photon and realized the quantum resetting process (Fig.~\ref{fig:setup}). More importantly, we demonstrated that the quantum correlation and the underlying form of entangled state is preserved when half of an entangled pair was subject to the reset. This was confirmed by violating the Clauser-Horne-Shimony-Holt (CHSH) inequality~\cite{Bell1964,chsh}. The preservation of entanglement shows that the protocols in~\cite{NavascuesPRX2018} can reset arbitrary quantum states, even beyond the capability of spin echo~\cite{PhysRev.80.580} or dynamical decoupling~~\cite{PhysRevLett.82.2417,zanardi}.

A schematic of the resetting protocol, implemented in our experiment, is shown in Fig.~\ref{fig:setup}(a). At time $t=0$, we initialize the target system $\mathcal{S}$ in an arbitrary state $\ket{\psi(0)}$ and prepare four probes $\mathcal{P}_1$, $\mathcal{P}_2$, $\mathcal{P}_3$, $\mathcal{P}_4$ in the quantum state $\ket{\Psi^{-}}_{\mathcal{P}_1\mathcal{P}_2}\otimes\ket{\Psi^{-}}_{\mathcal{P}_3\mathcal{P}_4}$, where $\ket{\Psi^{-}}=(1/\sqrt{2})(\ket{01}-\ket{10})$ is the singlet state and $\{\ket{0},\ket{1}\}$ is the computational basis. The target first evolves freely, following the unitary $U_0=e^{-i T\mathcal{H}_0}$, where we set $\hbar=1$. Then, the probe $\mathcal{P}_1$ interacts with the target via $V_{\mathcal{SP}}$ for a time $\delta$. At time $t=T+\delta$, the interaction with $\mathcal{P}_1$ ends and the target is again allowed to evolve freely for a time $T$, after which time it is probed by $\mathcal{P}_2$ for a time $\delta$. Similarly, the probe $\mathcal{P}_3$ ($\mathcal{P}_4$) interacts with the target at time $t=3T+2\delta$ ($t=4T+3\delta$) via the same gate $V_{\mathcal{SP}}$. At time $t=4T+4\delta$, all probes have interacted with the target and project them onto a quasi-symmetric space $Q$ spanned by six vectors,
{
\begin{equation}
\begin{aligned}
\ket{m_1}&=\ket{0000},\\
\ket{m_2}&=(\ket{0001}+\ket{0010}+\ket{0100}+\ket{1000})/2,\\
\ket{m_3}&=(\ket{0101}+\ket{0110}+\ket{1001}+\ket{1010})/2,\\
\ket{m_4}&=(\ket{0011}+\ket{1100})/\sqrt{2},\\
\ket{m_5}&=(\ket{0111}+\ket{1011}+\ket{1101}+\ket{1110})/2,\\
\ket{m_6}&=\ket{1111}.\\
\end{aligned}
\end{equation}
}
If we successfully project the probes to one of the six bases, the target will be brought back to the state $\ket{\psi(0)}$, otherwise the protocol fails.

We test the protocol with combinations of different $\mathcal{H}_0$, $U_0$ and $V_{\mathcal{SP}}$. The most general time-independent Hamiltonian describing the dynamics of a two-level quantum system such as our target is given by $\mathcal{H}_0=\textbf{n}\cdot\boldsymbol{\sigma}/2$, where $\textbf{n}=(\sin\theta\cos\phi,\sin\theta\sin\phi,\cos\theta)$ is a unit vector and $\boldsymbol{\sigma}=(\sigma_{x},\sigma_{y},\sigma_{z})$ is the Pauli vector. The evolution of the target can be seen as a $SU(2)$ rotation through an axis given by $\textbf{n}$. Here we primarily present the results by setting $\theta=0$. The free Hamiltonian is $\mathcal{H}_0=\sigma_{z}/2$, and the evolution  from time $t_1$ to $t_2$ can be described by
\begin{equation}
\begin{aligned}
U_0=\begin{pmatrix}
e^{-i(t_{2}-t_{1})/2} & 0\\
0 & e^{i(t_{2}-t_{1})/2}\\
\end{pmatrix}.
\end{aligned}
\end{equation}
The results for $\theta=\pi/2$ and $\phi=0$, i.e., $\mathcal{H}_0=\sigma_{y}/2$, are shown in the Supplementary Material.

As shown in ref.~\cite{NavascuesPRX2018}, the interaction gate $V_{\mathcal{SP}}$ can either be unitary or a simple quantum channel (simple quantum channels are non-unitary). We designed two optical circuits, $V_{\mathcal{SP}}^{\text{I}}$ and $V_{\mathcal{SP}}^{\text{II}}$, illustrated in Fig.~\ref{fig:setup}(b-c), to test each of these possibilities. For circuit I (Fig.~\ref{fig:setup}B), the interaction is unitary and is chosen as the SWAP gate, given by
\begin{equation}
\begin{aligned}
V_{\mathcal{SP}}^{\text{I}}=\text{SWAP}=\begin{pmatrix}
1 & 0 & 0 & 0\\
0 & 0 & 1 & 0\\
0 & 1 & 0 & 0\\
0 & 0 & 0 & 1\\
\end{pmatrix}.
\end{aligned}
\end{equation}

{Via calculation we find that with this $V_{\mathcal{SP}}^{\text{I}}$ the probes can only be projected onto $\ket{m_3}$ and $\ket{m_4}$ vectors of $Q$. For convenience but without loss of generality, we measure the probes in the state $\ket{m_3}$ by performing two Bell state measurements (BSM) on $\mathcal{P}_1,\mathcal{P}_2$ and $\mathcal{P}_3, \mathcal{P}_4$ simultaneously in the experiment.} The success of the protocol is heralded by the outputs of both BSMs being $\ket{\Psi^{+}}$, and the theoretical success probability is 1/16.

The interaction gate implemented by circuit II, is a non-unitary gate. That is, $V_{\mathcal{SP}}^{\text{II}}$ (Fig.~\ref{fig:setup}(c)), consists of a Pauli $X$ gate, a Hadamard gate ($H$), and the non-unitary gate $\text{G}_{\text{PBS}}$, implemented using a polarizing beam splitter (PBS). The whole gate is given by
\begin{equation}
V_{\mathcal{SP}}^{\text{II}}=(I\otimes H)\cdot\text{G}_{\text{PBS}}\cdot(X\otimes I)=\begin{pmatrix}
0 & 0 & \frac{1}{\sqrt{2}} & 0 \\
0 & 0 & \frac{1}{\sqrt{2}} & 0 \\
0 & \frac{1}{\sqrt{2}} & 0 & 0 \\
0 & -\frac{1}{\sqrt{2}} & 0 & 0 \\
\end{pmatrix}.
\end{equation}

Similar to $V_{\mathcal{SP}}^{\text{I}}$, the probes can only be projected onto $\ket{m_1}$, $\ket{m_4}$ and $\ket{m_6}$ vectors of $Q$. In the experiment, {we project the four probes into $\ket{m_1}$ and $\ket{m_6}$ by measuring each probe in the Pauli $Z$ basis.} We neglected $\ket{m_4}$ to simplify the experimental setup, causing the theoretical success probability to drop to 1/32. With all projections, when the interaction is sampled from the Haar measure, the average success probability of the  protocols is above 0.2 and there even exist interactions with a theoretical success probability of 1~\cite{NavascuesPRX2018}.

The target and the probes in our experiment are encoded by the polarization degree of freedom of photons~\cite{PanRMP2012}, with $\ket{0}$ corresponding to the horizontal polarization ($\ket{H}$) and $\ket{1}$ to vertical ($\ket{V}$) (Fig.~\ref{fig:setup}(d-e)). Three entangled photon pairs are generated by three independent nonlinear crystals through spontaneous parametric down-conversion (SPDC)~\cite{kwiat1995new}. To achieve this, a pulsed ultraviolet laser with central wavelength $390$ nm and pulse duration $140$ fs at a repetition rate of $76$ MHz {successively passes through three sandwich-like combinations of $\beta$-barium borate crystals (C-BBO).}  By the end of this process, three Einstein-Podolsky-Rosen (EPR) pairs have been generated. The first pair, $\ket{\Psi^{-}_{12}}$, where $\ket{\Psi^{-}}=(\ket{HV}-\ket{VH})/\sqrt{2}$, is used to create an arbitrary initial state $\ket{\psi_{0}}=\alpha\ket{H}+\beta\ket{V}$ in a heralded manner by projecting photon $1$ to $\beta^*\ket{H}-\alpha^*\ket{V}$. And the other two pairs, $\ket{\Psi^{-}_{34}}$ and $\ket{\Psi^{-}_{56}}$, serve as probing photons.

For circuit I (see Fig.~\ref{fig:setup}(d)), the SWAP gate is implemented by exchanging the photon channels. To realize two BSMs simultaneously, photons in channels 3\&4 and 5\&6 are overlapped at PBS$_1$ and PBS$_2$, and after overlapping, then a measurement in the $X$ basis is performed at each PBS output. The protocol succeeds when both of the BSMs give the result $\ket{\Psi^{+}}$. For circuit II (see Fig.~\ref{fig:setup}(e)), we implement $V_{\mathcal{SP}}^{\text{II}}$ with a PBS and two half-wave plates (HWPs). In the experiment, the target, encoded in photon 2, successively overlaps with the probe photons 3, 4, 5 and 6 at four PBSs. After overlapping, each probe is measured in the $\ket{H}/\ket{V}$ basis using a combination of quarter- and half-wave plates and a measurement PBS. When all four probes are detected at the transmissive (reflective) path of the measurement PBS, they are projected into basis $\ket{m_1}$ ($\ket{m_6}$) and thus the protocol succeeds.

\begin{figure}
	\centering
	\includegraphics[width=\linewidth]{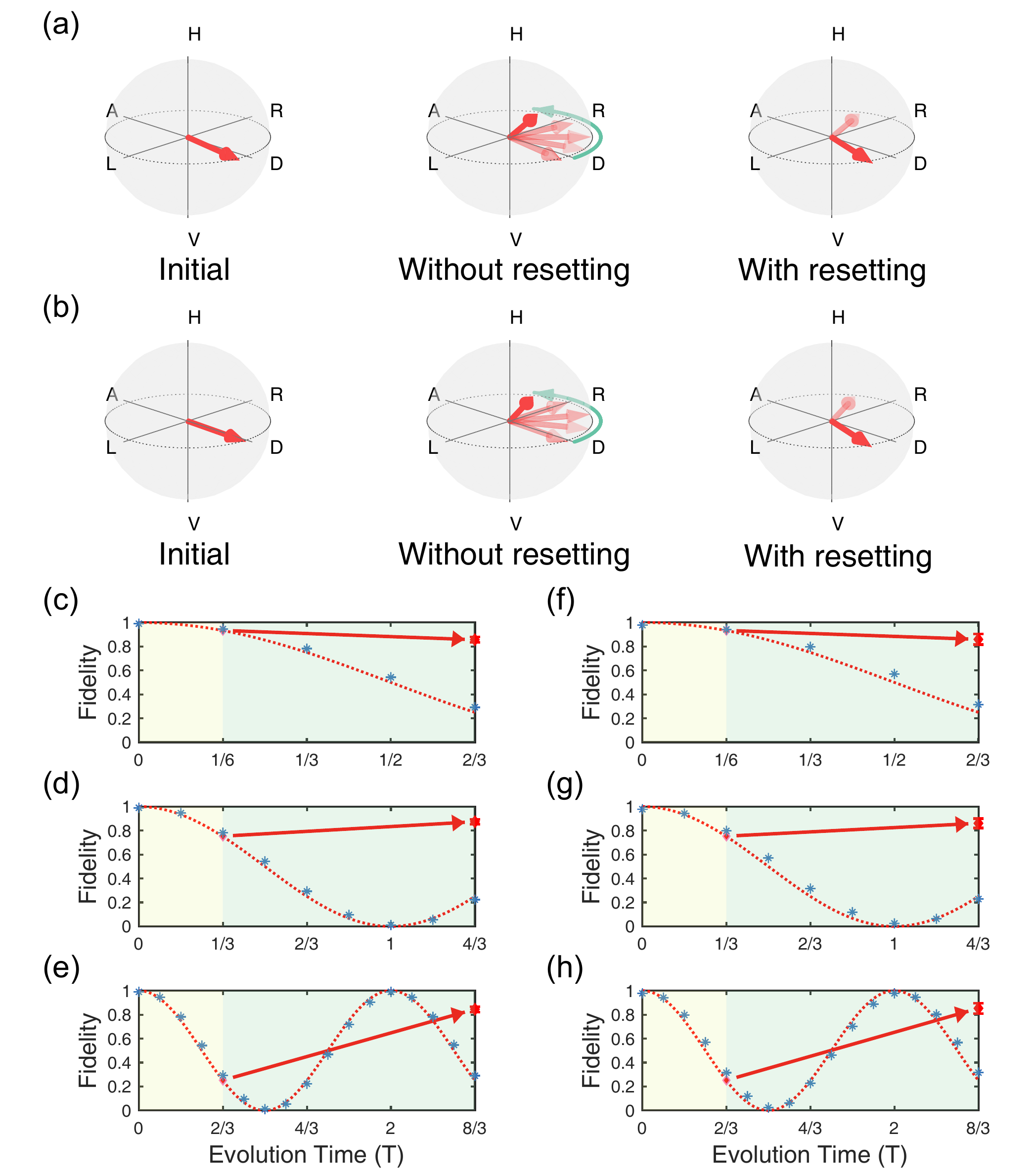}
	\caption{\textbf{Illustration of the quantum resetting process.} (a,b) Visual description of the evolution process before and after the resetting processing for circuit I and circuit II. (c-e) The evolution of state fidelity during resetting process for circuit I. (f-h) The evolution of state fidelity during resetting process for circuit II. The dashed curves and the blue asterisks denote, respectively, the ideal and the experimental fidelity with free time-dependent Hamiltonian evolution, without resetting process. The solid line and the red dots denote the experimental results of quantum resetting.
	}
	\label{fig:Evolution}
\end{figure}

In Fig.~\ref{fig:Evolution}, we demonstrate, visually and quantitatively, that the target quantum system can be reset after its uncontrolled dynamics have been neutralized. We initialize the target system at state $\ket{\psi(0)}=\ket{D}$ and let it evolve freely for different times: $T=\pi/6$, $T=\pi/3$, and $T=2\pi/3$. In Fig.~\ref{fig:Evolution}(a,b), we illustrate the process of free evolution on Bloch sphere with and without resetting for a visual description. With resetting, the target can be reset back to the initial state at time $t_0$ with high accuracy.
We quantify the effectiveness of the resetting process by defining the fidelity as the trace distance between the evolved state at time $t$ and the initial state at time $t_0=0$, $F=\text{Tr}(\rho_{t}\rho_{t_0})$. By also, evolutions with no resetting process are shown for reference.  The data points are in good agreement with the theoretical curve.
In all cases, the target quantum system can be successfully reset to its original state with a fidelity of at least 80\% (See Fig.\ref{fig:Evolution}(c-h)).


\begin{figure}
	\centering
	\includegraphics[width=\linewidth]{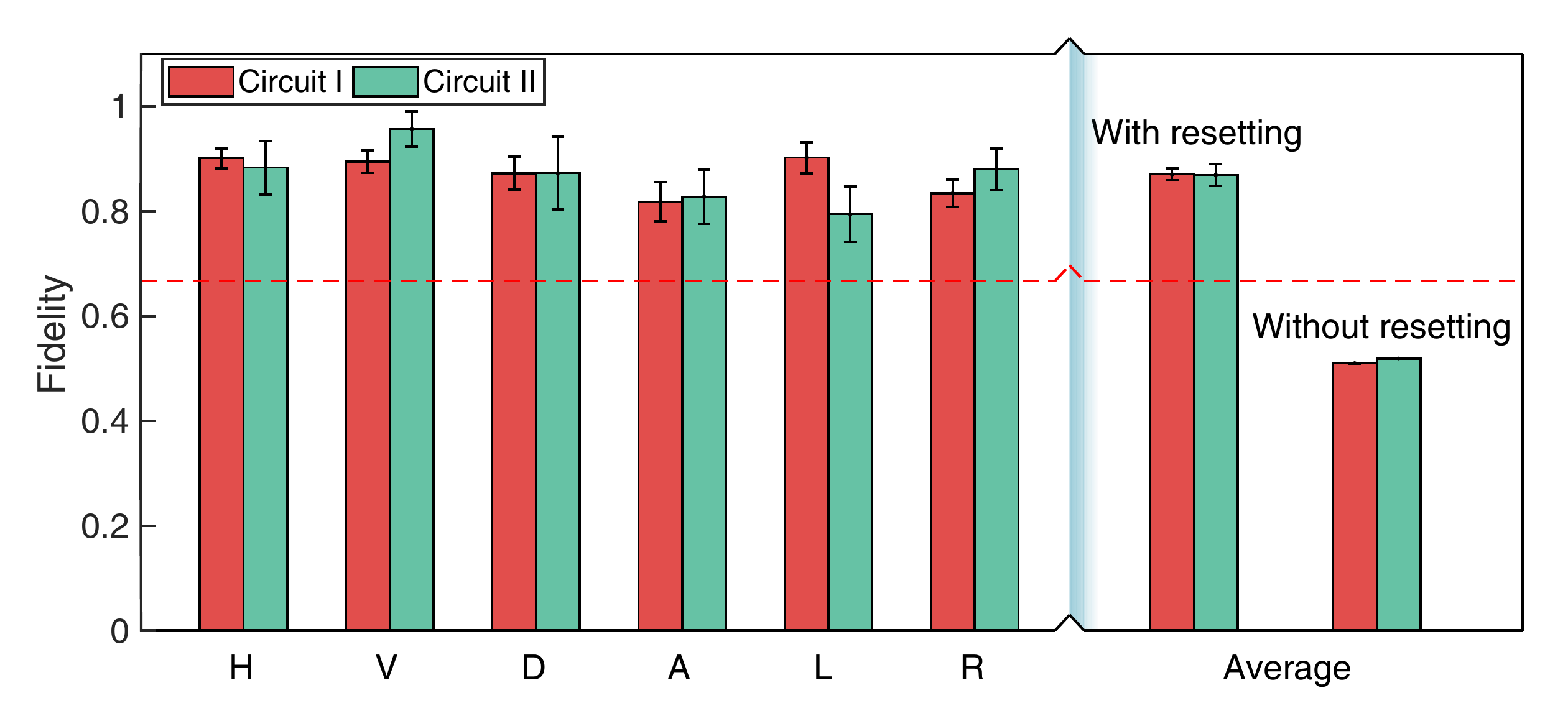}
	\caption{\textbf{The fidelity for different types of target states.} With resetting, all fidelities are well above the classical limit of 2/3 (indicated by the dashed curve), and the average fidelity for circuit I (II) is $F=0.870\pm0.012$ ($F=0.869\pm0.021$). Without resetting, the average fidelity for circuit I (II) is $F=0.5100\pm0.0005$ ($F=0.5186\pm0.0003$).
	}
	\label{fig:fidelity}
\end{figure}

To further verify the reliability of protocol in all cases, we initialized the target quantum system at different types of states and evaluated the fidelity between the reset state (after resetting) $\rho_{\text{reset}}$ and the ideal state $\rho_{\text{ideal}}$, $F=\text{Tr}(\rho_{\text{ideal}}\rho_{\text{reset}})$. For the two circuits, we prepared the target quantum system at six initial states, namely $\ket{H}$, $\ket{V}$, $\ket{D}$, $\ket{A}$, $\ket{L}$, and $\ket{R}$, where $\ket{D/A}=\ket{H}\pm\ket{V}$, $\ket{L/R}=\ket{H}\mp i\ket{V}$. We set the time of free evolution for each initial state to $T=2\pi/3$ for both circuit I and circuit II. With resetting, each of the six fidelities in both circuits is far beyond the classical limit of 2/3, and the average fidelities for circuit I and circuit II are $F=0.870\pm0.012$ and $F=0.869\pm0.021$, respectively (Fig.\ref{fig:fidelity}).
\begin{figure}
	\centering
	\includegraphics[width=\linewidth]{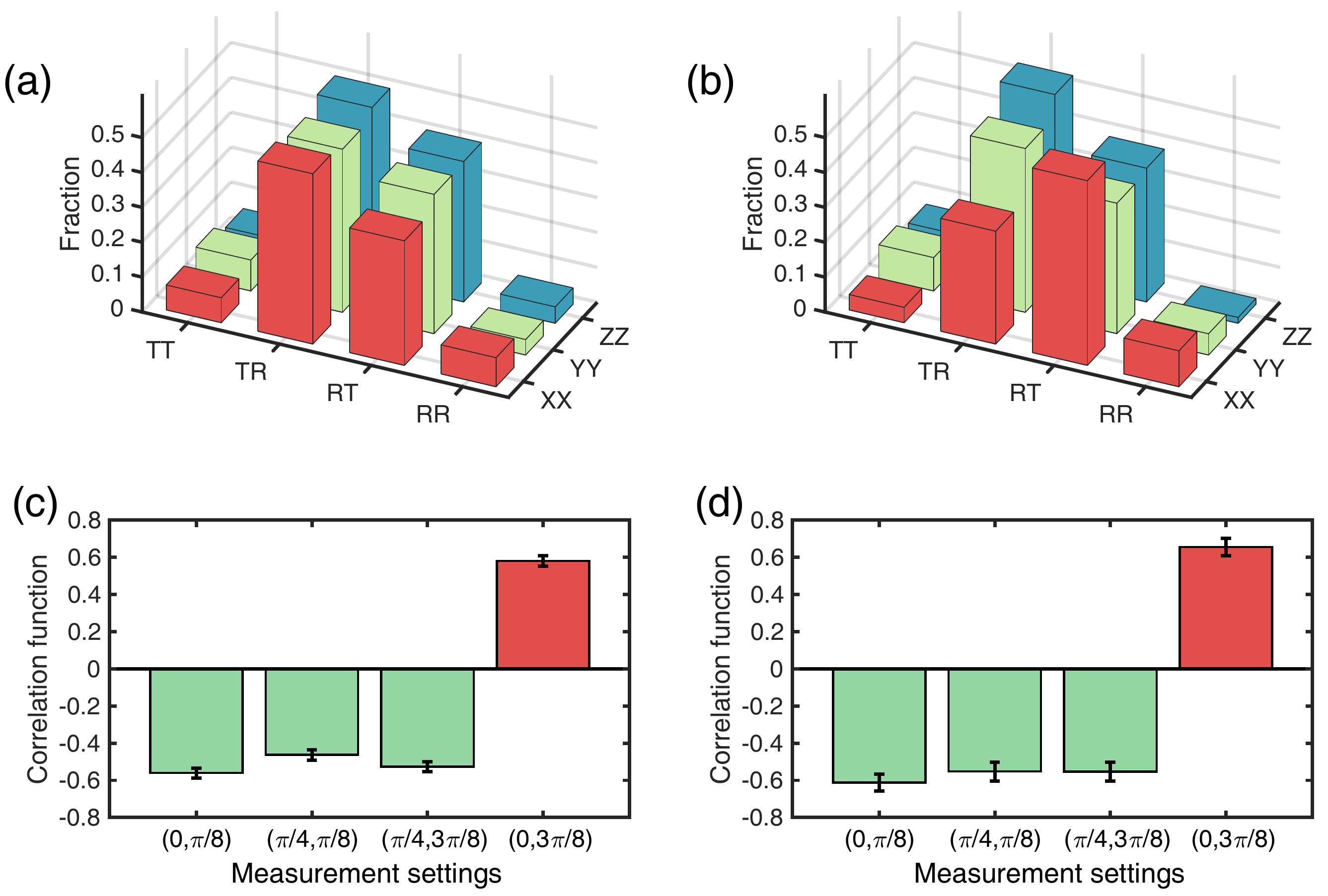}
	\caption{\textbf{The measured fraction and correlation of reset entanglement.} The measurement of entanglement fidelity and the CHSH inequality was conducted for entangled photon pairs, where one party of the pairs goes through the quantum resetting process. (a) The measurement was performed in basis $XX$, $YY$ and $ZZ$ for transmitted (T) and reflected (R) modes of the PBS. The entanglement fidelity for Circuit I is $F=0.805\pm0.017$. (b) The entanglement fidelity for Circuit II is $F=0.807\pm0.033$. (c) CHSH inequality value for Circuit I is $2.13\pm0.05$. (d) CHSH inequality value for Circuit II is $2.26\pm0.08$.
	}
	\label{fig:fraction}
\end{figure}

As a test of the quantum nature of the protocol, we show that when only part of an entangled state is reset, and the whole entangled state is preserved. As shown in Fig.~\ref{fig:setup}(d-e), photons 1 and 2 are prepared in the maximally entangled state $\ket{\Psi^{-}}$ as before. But only photon 2 is subjected to a nontrivial evolution and reset to its initial state. We can compute the entanglement fidelity between the state of photons 1, 2 before and after the protocol. The density matrix of $\ket{\Psi^{-}}$ can be decomposed as $\ket{\Psi^{-}}\bra{\Psi^{-}}=\frac{1}{4}(I-XX-YY-ZZ)$, where $Z=\ket{H}\bra{H}-\ket{V}\bra{V}$, $X=\ket{D}\bra{D}-\ket{A}\bra{A}$, and $Y=\ket{R}\bra{R}-\ket{L}\bra{L}$. Using such a decomposition, it is sufficient to measure in three bases, $XX$, $YY$, and $ZZ$, to obtain the entanglement fidelity. We set the free evolution time to be $T=2\pi/3$ for circuits I and II. We count the six-fold coincidence in 4 hours for each setting to obtain the measured fractions in Fig.~\ref{fig:fraction}(a-b). The entanglement fidelities are $F=0.805\pm0.017$ for circuit I and $F=0.807\pm0.033$ for circuit II. Furthermore, we perform a Bell test using the CHSH inequality with photons 1 and 2 in which only photon 2 undergoes the quantum resetting process. The results are shown in Fig.~\ref{fig:fraction}(c-d). We obtain a CHSH value of $2.13\pm0.05$ ($2.26\pm0.08$) for circuit I (II). Without resetting, the CHSH value is about $0.710\pm0.002$ under the measurement settings of Fig.~\ref{fig:fraction}(c-d). The Bell violation shows that the quantum resetting protocol preserves the entanglement, and more importantly the entangled state, in a composite quantum system when it is applied to part of it.

In conclusion, we have experimentally demonstrated that the free evolution of an uncontrolled two-level quantum system, such as a photon polarized in the horizontal and vertical directions, can be turned back and undone. This is accomplished probabilistically by making it interact, not necessarily unitarily, with four quantum probes. This process effectively alters the flow of time experienced by the uncontrolled system. Also, we show that when this alteration happens to one part of a composite quantum system, the correlation present in it is preserved at the end of the process.



{An extension of the resetting protocol also allows the dynamics of multiple qubit targets to be treated as a resource: by slowing down or freezing some parts of a system, other parts can be accelerated~\cite{navascues_new}. However, implementing this protocol requires significantly more probes and more complex interactions. The scaling of the probes is also the main hindrance in implementing the resetting protocol for higher dimensional targets. Fortunately, recent progress in both theoretical~\cite{CiracZollerOptLattBosonFermion} and experimental~\cite{GrossBlochOptLattReview} techniques has shown ultracold atoms in optical lattices to be a promising candidate for simulating complex many-body quantum systems. Photons cool down the atoms then trap them in a periodic potential. Interactions between the atoms can be controlled by tuning the shape of the trap. It is possible to incorporate our photonic implementation of the resetting protocol into such lattices to help us overcome the challenges, by using photonic systems as the control mechanism and atomic systems as the targets.}

This work was supported by the National Key R\&D Program of China (under Grants No. 2018YFB0504303 and No. 2018YFA0306501), the National Natural Science Foundation of China (under Grants No. 11425417, No. 61771443 and U1738140), Fundamental Research Funds for the Central Universities (WK2340000083), the Anhui Initiative in Quantum Information Technologies and the Chinese Academy of Sciences. Z.W. was supported by the National Key R\&D Program of China (under Grant No. 2018YFA0306703).


\begin{thebibliography}{17}%
\makeatletter
\providecommand \@ifxundefined [1]{%
 \@ifx{#1\undefined}
}%
\providecommand \@ifnum [1]{%
 \ifnum #1\expandafter \@firstoftwo
 \else \expandafter \@secondoftwo
 \fi
}%
\providecommand \@ifx [1]{%
 \ifx #1\expandafter \@firstoftwo
 \else \expandafter \@secondoftwo
 \fi
}%
\providecommand \natexlab [1]{#1}%
\providecommand \enquote  [1]{``#1''}%
\providecommand \bibnamefont  [1]{#1}%
\providecommand \bibfnamefont [1]{#1}%
\providecommand \citenamefont [1]{#1}%
\providecommand \href@noop [0]{\@secondoftwo}%
\providecommand \href [0]{\begingroup \@sanitize@url \@href}%
\providecommand \@href[1]{\@@startlink{#1}\@@href}%
\providecommand \@@href[1]{\endgroup#1\@@endlink}%
\providecommand \@sanitize@url [0]{\catcode `\\12\catcode `\$12\catcode
  `\&12\catcode `\#12\catcode `\^12\catcode `\_12\catcode `\%12\relax}%
\providecommand \@@startlink[1]{}%
\providecommand \@@endlink[0]{}%
\providecommand \url  [0]{\begingroup\@sanitize@url \@url }%
\providecommand \@url [1]{\endgroup\@href {#1}{\urlprefix }}%
\providecommand \urlprefix  [0]{URL }%
\providecommand \Eprint [0]{\href }%
\providecommand \doibase [0]{https://doi.org/}%
\providecommand \selectlanguage [0]{\@gobble}%
\providecommand \bibinfo  [0]{\@secondoftwo}%
\providecommand \bibfield  [0]{\@secondoftwo}%
\providecommand \translation [1]{[#1]}%
\providecommand \BibitemOpen [0]{}%
\providecommand \bibitemStop [0]{}%
\providecommand \bibitemNoStop [0]{.\EOS\space}%
\providecommand \EOS [0]{\spacefactor3000\relax}%
\providecommand \BibitemShut  [1]{\csname bibitem#1\endcsname}%
\let\auto@bib@innerbib\@empty
\bibitem [{\citenamefont {Rabitz}\ \emph {et~al.}(2000)\citenamefont {Rabitz},
  \citenamefont {de~Vivie-Riedle}, \citenamefont {Motzkus},\ and\ \citenamefont
  {Kompa}}]{RabitzQuantumControl}%
  \BibitemOpen
  \bibfield  {author} {\bibinfo {author} {\bibfnamefont {H.}~\bibnamefont
  {Rabitz}}, \bibinfo {author} {\bibfnamefont {R.}~\bibnamefont
  {de~Vivie-Riedle}}, \bibinfo {author} {\bibfnamefont {M.}~\bibnamefont
  {Motzkus}},\ and\ \bibinfo {author} {\bibfnamefont {K.}~\bibnamefont
  {Kompa}},\ }\bibfield  {title} {\bibinfo {title} {Whither the future of
  controlling quantum phenomena?},\ }\href
  {https://doi.org/10.1126/science.288.5467.824} {\bibfield  {journal}
  {\bibinfo  {journal} {Science}\ }\textbf {\bibinfo {volume} {288}},\ \bibinfo
  {pages} {824} (\bibinfo {year} {2000})}\BibitemShut {NoStop}%
\bibitem [{\citenamefont {Hahn}(1950)}]{PhysRev.80.580}%
  \BibitemOpen
  \bibfield  {author} {\bibinfo {author} {\bibfnamefont {E.~L.}\ \bibnamefont
  {Hahn}},\ }\bibfield  {title} {\bibinfo {title} {{Spin Echoes}},\ }\href
  {https://doi.org/10.1103/PhysRev.80.580} {\bibfield  {journal} {\bibinfo
  {journal} {Phys. Rev.}\ }\textbf {\bibinfo {volume} {80}},\ \bibinfo {pages}
  {580} (\bibinfo {year} {1950})}\BibitemShut {NoStop}%
\bibitem [{\citenamefont {Freeman}(1998)}]{freeman1998spin}%
  \BibitemOpen
  \bibfield  {author} {\bibinfo {author} {\bibfnamefont {R.}~\bibnamefont
  {Freeman}},\ }\href@noop {} {\emph {\bibinfo {title} {{Spin Choreography:
  Basic Steps in High Resolution NMR}}}}\ (\bibinfo  {publisher} {Oxford
  University Press},\ \bibinfo {year} {1998})\BibitemShut {NoStop}%
\bibitem [{\citenamefont {Viola}\ \emph {et~al.}(1999)\citenamefont {Viola},
  \citenamefont {Knill},\ and\ \citenamefont {Lloyd}}]{PhysRevLett.82.2417}%
  \BibitemOpen
  \bibfield  {author} {\bibinfo {author} {\bibfnamefont {L.}~\bibnamefont
  {Viola}}, \bibinfo {author} {\bibfnamefont {E.}~\bibnamefont {Knill}},\ and\
  \bibinfo {author} {\bibfnamefont {S.}~\bibnamefont {Lloyd}},\ }\bibfield
  {title} {\bibinfo {title} {Dynamical decoupling of open quantum systems},\
  }\href {https://doi.org/10.1103/PhysRevLett.82.2417} {\bibfield  {journal}
  {\bibinfo  {journal} {Phys. Rev. Lett.}\ }\textbf {\bibinfo {volume} {82}},\
  \bibinfo {pages} {2417} (\bibinfo {year} {1999})}\BibitemShut {NoStop}%
\bibitem [{\citenamefont {Zanardi}(1999)}]{zanardi}%
  \BibitemOpen
  \bibfield  {author} {\bibinfo {author} {\bibfnamefont {P.}~\bibnamefont
  {Zanardi}},\ }\bibfield  {title} {\bibinfo {title} {{Symmetrizing
  evolutions}},\ }\href {https://doi.org/10.1016/S0375-9601(99)00365-5}
  {\bibfield  {journal} {\bibinfo  {journal} {Phys. Lett. A}\ }\textbf
  {\bibinfo {volume} {258}},\ \bibinfo {pages} {77 } (\bibinfo {year}
  {1999})}\BibitemShut {NoStop}%
\bibitem [{\citenamefont {Sardharwalla}\ \emph {et~al.}(2016)\citenamefont
  {Sardharwalla}, \citenamefont {Cubitt}, \citenamefont {Harrow},\ and\
  \citenamefont {Linden}}]{1602.07963}%
  \BibitemOpen
  \bibfield  {author} {\bibinfo {author} {\bibfnamefont {I.~S.~B.}\
  \bibnamefont {Sardharwalla}}, \bibinfo {author} {\bibfnamefont {T.~S.}\
  \bibnamefont {Cubitt}}, \bibinfo {author} {\bibfnamefont {A.~W.}\
  \bibnamefont {Harrow}},\ and\ \bibinfo {author} {\bibfnamefont
  {N.}~\bibnamefont {Linden}},\ }\href@noop {} {\emph {\bibinfo {title}
  {{Universal Refocusing of Systematic Quantum Noise}}}},\ \bibinfo {type}
  {Tech. Rep.}\ \bibinfo {number} {MIT/CTP-4772}\ (\bibinfo  {institution}
  {MIT},\ \bibinfo {year} {2016})\ \Eprint
  {https://arxiv.org/abs/arXiv:1602.07963} {arXiv:1602.07963} \BibitemShut
  {NoStop}%
\bibitem [{\citenamefont {Aharonov}\ \emph {et~al.}(1990)\citenamefont
  {Aharonov}, \citenamefont {Anandan}, \citenamefont {Popescu},\ and\
  \citenamefont {Vaidman}}]{PhysRevLett.64.2965}%
  \BibitemOpen
  \bibfield  {author} {\bibinfo {author} {\bibfnamefont {Y.}~\bibnamefont
  {Aharonov}}, \bibinfo {author} {\bibfnamefont {J.}~\bibnamefont {Anandan}},
  \bibinfo {author} {\bibfnamefont {S.}~\bibnamefont {Popescu}},\ and\ \bibinfo
  {author} {\bibfnamefont {L.}~\bibnamefont {Vaidman}},\ }\bibfield  {title}
  {\bibinfo {title} {{Superpositions of time evolutions of a quantum system and
  a quantum time-translation machine}},\ }\href
  {https://doi.org/10.1103/PhysRevLett.64.2965} {\bibfield  {journal} {\bibinfo
   {journal} {Phys. Rev. Lett.}\ }\textbf {\bibinfo {volume} {64}},\ \bibinfo
  {pages} {2965} (\bibinfo {year} {1990})}\BibitemShut {NoStop}%
\bibitem [{\citenamefont {Drensky}\ and\ \citenamefont
  {Formanek}(2004)}]{drensky_formanek_book}%
  \BibitemOpen
  \bibfield  {author} {\bibinfo {author} {\bibfnamefont {V.}~\bibnamefont
  {Drensky}}\ and\ \bibinfo {author} {\bibfnamefont {E.}~\bibnamefont
  {Formanek}},\ }\href@noop {} {\emph {\bibinfo {title} {Polynomial identity
  rings}}},\ Advanced Courses in Mathematics CRM Barcelona\ (\bibinfo
  {publisher} {Birkh{\"a}user},\ \bibinfo {year} {2004})\BibitemShut {NoStop}%
\bibitem [{\citenamefont {Navascu\'es}(2018)}]{NavascuesPRX2018}%
  \BibitemOpen
  \bibfield  {author} {\bibinfo {author} {\bibfnamefont {M.}~\bibnamefont
  {Navascu\'es}},\ }\bibfield  {title} {\bibinfo {title} {Resetting
  uncontrolled quantum systems},\ }\href
  {https://doi.org/10.1103/PhysRevX.8.031008} {\bibfield  {journal} {\bibinfo
  {journal} {Phys. Rev. X}\ }\textbf {\bibinfo {volume} {8}},\ \bibinfo {pages}
  {031008} (\bibinfo {year} {2018})}\BibitemShut {NoStop}%
\bibitem [{\citenamefont {Quintino}\ \emph {et~al.}(2019)\citenamefont
  {Quintino}, \citenamefont {Dong}, \citenamefont {Shimbo}, \citenamefont
  {Soeda},\ and\ \citenamefont {Murao}}]{PhysRevLett.123.210502}%
  \BibitemOpen
  \bibfield  {author} {\bibinfo {author} {\bibfnamefont {M.~T.}\ \bibnamefont
  {Quintino}}, \bibinfo {author} {\bibfnamefont {Q.}~\bibnamefont {Dong}},
  \bibinfo {author} {\bibfnamefont {A.}~\bibnamefont {Shimbo}}, \bibinfo
  {author} {\bibfnamefont {A.}~\bibnamefont {Soeda}},\ and\ \bibinfo {author}
  {\bibfnamefont {M.}~\bibnamefont {Murao}},\ }\bibfield  {title} {\bibinfo
  {title} {Reversing unknown quantum transformations: Universal quantum circuit
  for inverting general unitary operations},\ }\href
  {https://doi.org/10.1103/PhysRevLett.123.210502} {\bibfield  {journal}
  {\bibinfo  {journal} {Phys. Rev. Lett.}\ }\textbf {\bibinfo {volume} {123}},\
  \bibinfo {pages} {210502} (\bibinfo {year} {2019})}\BibitemShut {NoStop}%
\bibitem [{\citenamefont {Bell}(1964)}]{Bell1964}%
  \BibitemOpen
  \bibfield  {author} {\bibinfo {author} {\bibfnamefont {J.~S.}\ \bibnamefont
  {Bell}},\ }\bibfield  {title} {\bibinfo {title} {On the
  {Einstein-Podolsky-Rosen} paradox},\ }\href
  {https://doi.org/10.1103/PhysicsPhysiqueFizika.1.195} {\bibfield  {journal}
  {\bibinfo  {journal} {Physics}\ }\textbf {\bibinfo {volume} {1}},\ \bibinfo
  {pages} {195} (\bibinfo {year} {1964})}\BibitemShut {NoStop}%
\bibitem [{\citenamefont {Clauser}\ \emph {et~al.}(1969)\citenamefont
  {Clauser}, \citenamefont {Horne}, \citenamefont {Shimony},\ and\
  \citenamefont {Holt}}]{chsh}%
  \BibitemOpen
  \bibfield  {author} {\bibinfo {author} {\bibfnamefont {J.~F.}\ \bibnamefont
  {Clauser}}, \bibinfo {author} {\bibfnamefont {M.~A.}\ \bibnamefont {Horne}},
  \bibinfo {author} {\bibfnamefont {A.}~\bibnamefont {Shimony}},\ and\ \bibinfo
  {author} {\bibfnamefont {R.~A.}\ \bibnamefont {Holt}},\ }\bibfield  {title}
  {\bibinfo {title} {Proposed experiment to test local hidden-variable
  theories},\ }\href {https://doi.org/10.1103/PhysRevLett.23.880} {\bibfield
  {journal} {\bibinfo  {journal} {Phys. Rev. Lett.}\ }\textbf {\bibinfo
  {volume} {23}},\ \bibinfo {pages} {880} (\bibinfo {year} {1969})}\BibitemShut
  {NoStop}%
\bibitem [{\citenamefont {Pan}\ \emph {et~al.}(2012)\citenamefont {Pan},
  \citenamefont {Chen}, \citenamefont {Lu}, \citenamefont {Weinfurter},
  \citenamefont {Zeilinger},\ and\ \citenamefont {\ifmmode~\dot{Z}\else
  \.{Z}\fi{}ukowski}}]{PanRMP2012}%
  \BibitemOpen
  \bibfield  {author} {\bibinfo {author} {\bibfnamefont {J.-W.}\ \bibnamefont
  {Pan}}, \bibinfo {author} {\bibfnamefont {Z.-B.}\ \bibnamefont {Chen}},
  \bibinfo {author} {\bibfnamefont {C.-Y.}\ \bibnamefont {Lu}}, \bibinfo
  {author} {\bibfnamefont {H.}~\bibnamefont {Weinfurter}}, \bibinfo {author}
  {\bibfnamefont {A.}~\bibnamefont {Zeilinger}},\ and\ \bibinfo {author}
  {\bibfnamefont {M.}~\bibnamefont {\ifmmode~\dot{Z}\else \.{Z}\fi{}ukowski}},\
  }\bibfield  {title} {\bibinfo {title} {Multiphoton entanglement and
  interferometry},\ }\href {https://doi.org/10.1103/RevModPhys.84.777}
  {\bibfield  {journal} {\bibinfo  {journal} {Rev. Mod. Phys.}\ }\textbf
  {\bibinfo {volume} {84}},\ \bibinfo {pages} {777} (\bibinfo {year}
  {2012})}\BibitemShut {NoStop}%
\bibitem [{\citenamefont {Kwiat}\ \emph {et~al.}(1995)\citenamefont {Kwiat},
  \citenamefont {Mattle}, \citenamefont {Weinfurter}, \citenamefont
  {Zeilinger}, \citenamefont {Sergienko},\ and\ \citenamefont
  {Shih}}]{kwiat1995new}%
  \BibitemOpen
  \bibfield  {author} {\bibinfo {author} {\bibfnamefont {P.~G.}\ \bibnamefont
  {Kwiat}}, \bibinfo {author} {\bibfnamefont {K.}~\bibnamefont {Mattle}},
  \bibinfo {author} {\bibfnamefont {H.}~\bibnamefont {Weinfurter}}, \bibinfo
  {author} {\bibfnamefont {A.}~\bibnamefont {Zeilinger}}, \bibinfo {author}
  {\bibfnamefont {A.~V.}\ \bibnamefont {Sergienko}},\ and\ \bibinfo {author}
  {\bibfnamefont {Y.}~\bibnamefont {Shih}},\ }\bibfield  {title} {\bibinfo
  {title} {New high-intensity source of polarization-entangled photon pairs},\
  }\href {https://doi.org/10.1103/PhysRevLett.75.4337} {\bibfield  {journal}
  {\bibinfo  {journal} {Phys. Rev. Lett.}\ }\textbf {\bibinfo {volume} {75}},\
  \bibinfo {pages} {4337} (\bibinfo {year} {1995})}\BibitemShut {NoStop}%
\bibitem [{\citenamefont {Trillo}\ \emph {et~al.}()\citenamefont {Trillo},
  \citenamefont {Dive},\ and\ \citenamefont {Navascu\'{e}s}}]{navascues_new}%
  \BibitemOpen
  \bibfield  {author} {\bibinfo {author} {\bibfnamefont {D.}~\bibnamefont
  {Trillo}}, \bibinfo {author} {\bibfnamefont {B.}~\bibnamefont {Dive}},\ and\
  \bibinfo {author} {\bibfnamefont {M.}~\bibnamefont {Navascu\'{e}s}},\
  }\bibfield  {title} {\bibinfo {title} {Remote time manipulation},\ }\href
  {https://arxiv.org/abs/1903.10568} {\bibfield  {journal} {\bibinfo  {journal}
  {arXiv}\ }}\Eprint {https://arxiv.org/abs/1903.10568} {1903.10568}
  \BibitemShut {NoStop}%
\bibitem [{\citenamefont {Arg{\"u}ello-Luengo}\ \emph
  {et~al.}(2019)\citenamefont {Arg{\"u}ello-Luengo}, \citenamefont
  {Gonz{\'a}lez-Tudela}, \citenamefont {Shi}, \citenamefont {Zoller},\ and\
  \citenamefont {Cirac}}]{CiracZollerOptLattBosonFermion}%
  \BibitemOpen
  \bibfield  {author} {\bibinfo {author} {\bibfnamefont {J.}~\bibnamefont
  {Arg{\"u}ello-Luengo}}, \bibinfo {author} {\bibfnamefont {A.}~\bibnamefont
  {Gonz{\'a}lez-Tudela}}, \bibinfo {author} {\bibfnamefont {T.}~\bibnamefont
  {Shi}}, \bibinfo {author} {\bibfnamefont {P.}~\bibnamefont {Zoller}},\ and\
  \bibinfo {author} {\bibfnamefont {J.~I.}\ \bibnamefont {Cirac}},\ }\bibfield
  {title} {\bibinfo {title} {Analogue quantum chemistry simulation},\ }\href
  {https://doi.org/10.1038/s41586-019-1614-4} {\bibfield  {journal} {\bibinfo
  {journal} {Nature}\ }\textbf {\bibinfo {volume} {574}},\ \bibinfo {pages}
  {215} (\bibinfo {year} {2019})}\BibitemShut {NoStop}%
\bibitem [{\citenamefont {Gross}\ and\ \citenamefont
  {Bloch}(2017)}]{GrossBlochOptLattReview}%
  \BibitemOpen
  \bibfield  {author} {\bibinfo {author} {\bibfnamefont {C.}~\bibnamefont
  {Gross}}\ and\ \bibinfo {author} {\bibfnamefont {I.}~\bibnamefont {Bloch}},\
  }\bibfield  {title} {\bibinfo {title} {Quantum simulations with ultracold
  atoms in optical lattices},\ }\href {https://doi.org/10.1126/science.aal3837}
  {\bibfield  {journal} {\bibinfo  {journal} {Science}\ }\textbf {\bibinfo
  {volume} {357}},\ \bibinfo {pages} {995} (\bibinfo {year}
  {2017})}\BibitemShut {NoStop}%
\end{thebibliography}

%

\newpage

\appendix

\section{Experimental setup}

\begin{figure*}
	\centering
	\includegraphics[width=0.9\linewidth]{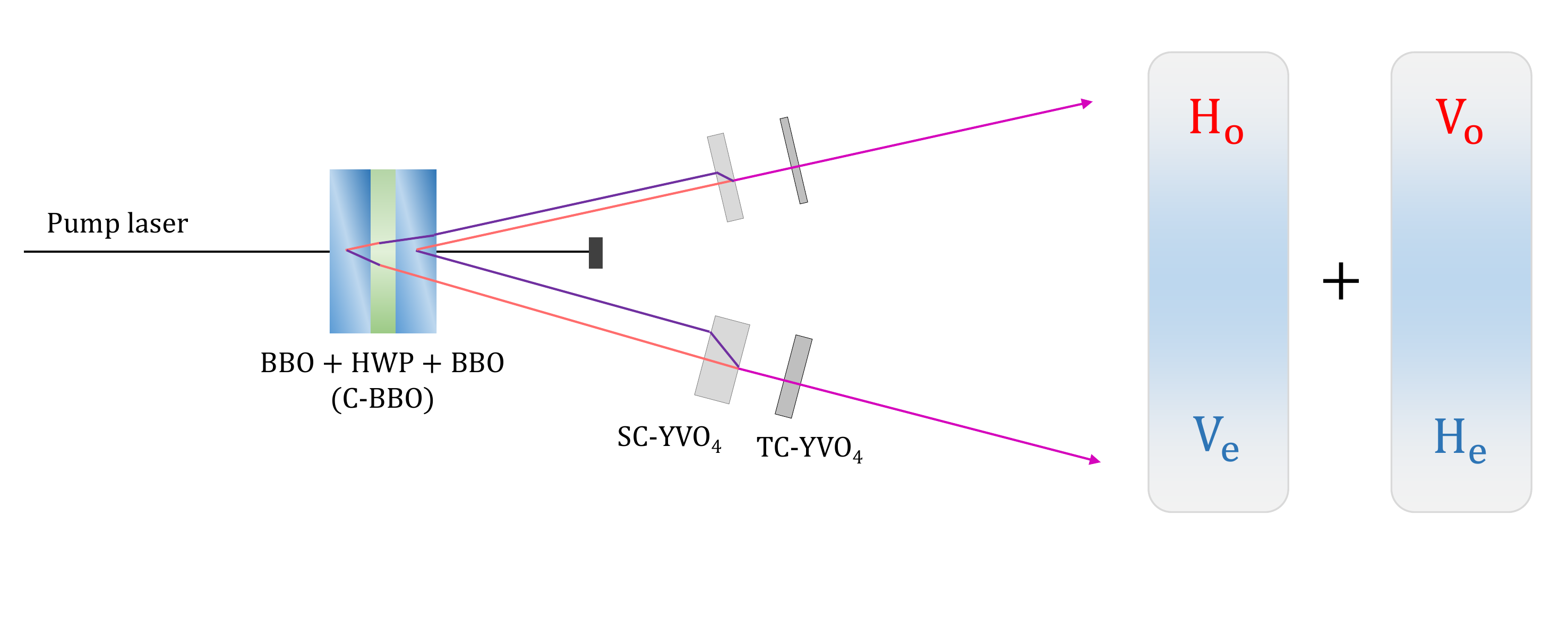}
	\caption{\textbf{Sandwich-like BBO+HWP+BBO geometry for generating entangled photons}
	}
	\label{fig:S4}
\end{figure*}

As shown in Fig. 1 in main text, for each entangled photon pair source, a $300$~mW pulsed ultraviolet laser is focused onto the sandwich-like geometry BBO crystal, which consists of two 2-mm-thick BBO crystals and a 44-$\mu$m-thick true-zero-order HWP, with the same waist $\omega_0\simeq$ 150 $\mu$m. Further, we use a YVO$_4$ crystal with a thickness of 2.3 (0.715) mm cut at $45^\circ$ with respective to the propagation direction in the horizontal plane to compensate spatial walk-off in the arm of extraordinary (ordinary) ray, and a 1.36 (0.39) mm thick YVO$_4$ cut at $90^\circ$ to compensate temporal walk-off (see Fig. \ref{fig:S4}). To eliminate the frequency correlations between independent pairs, we select bandpass filters with $3.6$ nm to spectrally filter both signal and idler photons. With this filter setting, the respective twofold coincidence court rate of three entangled photon pairs are $\sim$ 135000 Hz, $\sim$ 156000 Hz, $\sim$ 102000 Hz, correspond overall efficiency is $23.7\%$, $24.0\%$, $23.6\%$ and visibility in $\ket{D}/\ket{A}$ basis is $95.2\%$, $96.7\%$, $97.0\%$. With finely adjusting the distance of each photon, we make sure that the photons overlap on the PBSs well for both two circuit, and we eventually obtain an average fourfold coincidence rate of $\sim$21~$\text{s}^{-1}$ with a corresponding Hong-Ou-Mandel visibility of $89.5\%$ in $\ket{D}/\ket{A}$ basis.

\section{Noise analysis}

\begin{figure*}
	\centering
	\includegraphics[width=0.9\linewidth]{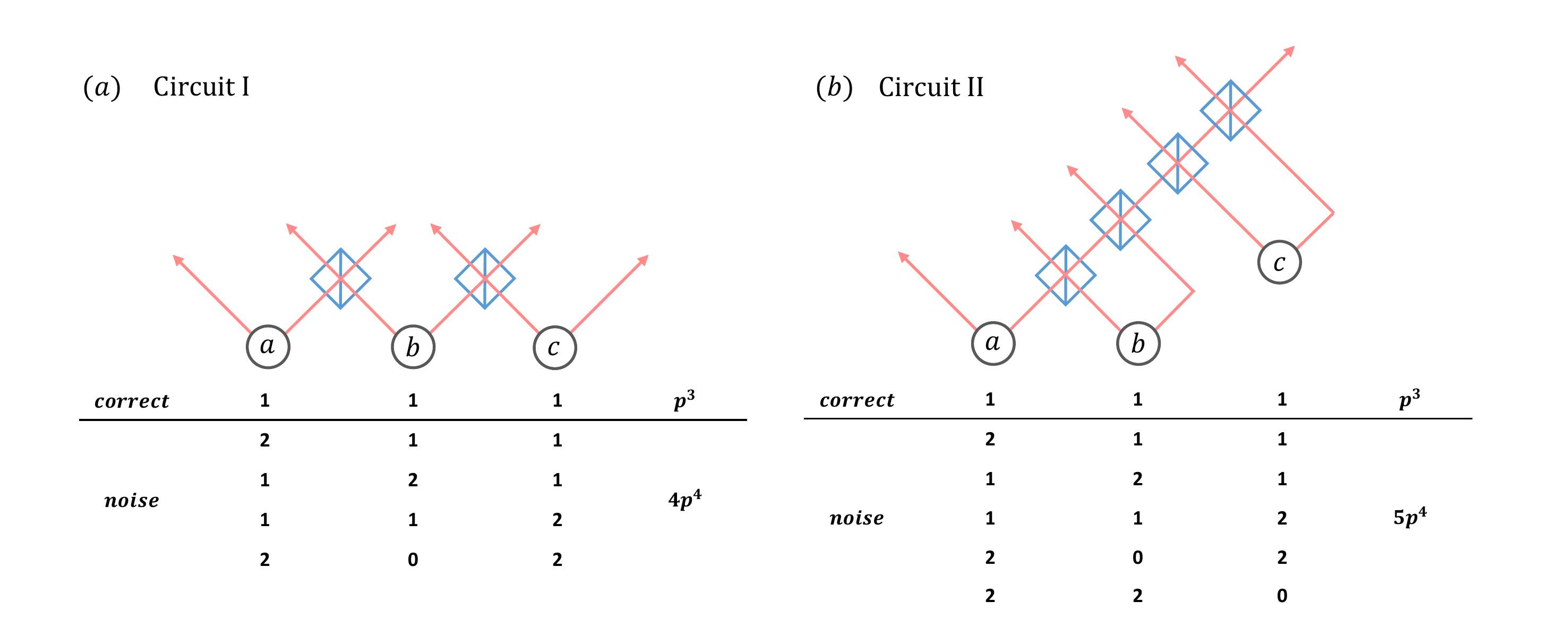}
	\caption{\textbf{Noise contribution from double-pair emission.}
	}
	\label{fig:S5}
\end{figure*}

{An important experimental challenge in the resetting process is noise control. The noise of our setup as shown in main text is mainly from the higher order spontaneous parametric down-conversion (SPDC) process and the temporal distinguishability between different photon pairs.
	
Here, we primarily analyze the noise from the higher order SPDC process in our experiment, include circuit $\uppercase\expandafter{\romannumeral1}$ and circuit $\uppercase\expandafter{\romannumeral2}$. In general, we define the $p$ as the downconversion probability, $M$ as the repetition frequency of the pulsed ultraviolet laser, $\eta$ as the overall collection efficiency which combine with the link, coupling and detection efficiency. Therefore, we can obtain $p_1 = 0.0316$, $p_2 = 0.0356$, $p_3 = 0.0241$ based on coincidence count rate $C = p \cdot M \cdot \eta$.

During quantum resetting protocol, we assume the effect of high-order emission on target system qubit $s$ and qubit $i$ which entangled with the target system can be regard as white noise model with probability $(1-\nu)$, under this scenario, the single photon state and two photons entangled state after resetting can be described as a Werner-like state:

\begin{equation}
\begin{aligned}
\rho_{reset}&=\nu \rho_{ideal} + (1-\nu)\cdot \frac{I^{\otimes 1}}{2},\\
\rho_{reset,~Ent}&=\nu \rho_{ideal,~Ent} + (1-\nu)\cdot \frac{I^{\otimes 2}}{4}.\\
\end{aligned}
\end{equation}

We consider the noisy contribution from our two circuits (see Fig. \ref{fig:S5}), for circuit $\uppercase\expandafter{\romannumeral1}$ and circuit $\uppercase\expandafter{\romannumeral2}$,

\begin{equation}
\begin{aligned}
\nu^{I}&=\frac{p^3}{p^3 + 4p^4},\\
\nu^{II}&=\frac{p^3}{p^3 + 5p^4}.\\
\end{aligned}
\end{equation}

According to the average downconversion probability $\overline{p} = 0.0304$, the theoretical fidelity of target system $\rho_s$ of circuit $\uppercase\expandafter{\romannumeral1}$ and circuit $\uppercase\expandafter{\romannumeral2}$ can be calculated as

\begin{equation}
\begin{aligned}
F^{\uppercase\expandafter{\romannumeral1}}&=\trace({\rho_{ideal}^{\uppercase\expandafter{\romannumeral1}}} \cdot {\rho_{reset}^{\uppercase\expandafter{\romannumeral1}}}) = 0.9458,\\
F^{\uppercase\expandafter{\romannumeral2}}&=\trace({\rho_{ideal}^{\uppercase\expandafter{\romannumeral2}}} \cdot {\rho_{reset}^{\uppercase\expandafter{\romannumeral2}}}) = 0.9340.\\
\end{aligned}
\end{equation}

The theoretical fidelity of two-photon entangled state $\rho_{si}$ of circuit $\uppercase\expandafter{\romannumeral1}$ and circuit $\uppercase\expandafter{\romannumeral2}$ can be calculated as

\begin{equation}
\begin{aligned}
F^{\uppercase\expandafter{\romannumeral1}}_{Ent}&=\trace({\rho_{ideal,~Ent}^{\uppercase\expandafter{\romannumeral1}}} \cdot {\rho_{reset,~Ent}^{\uppercase\expandafter{\romannumeral1}}}) = 0.9187,\\
F^{\uppercase\expandafter{\romannumeral2}}_{Ent}&=\trace({\rho_{ideal,~Ent}^{\uppercase\expandafter{\romannumeral2}}} \cdot {\rho_{reset,~Ent}^{\uppercase\expandafter{\romannumeral2}}}) = 0.9010.\\
\end{aligned}
\end{equation}

Moreover, the temporal distinguishability between different photon pairs, dark count of single photon detectors, white light noise in laboratory and the imperfections of optical elements (PBS, HWP, etc.) also may lead to the loss of fidelity.
In the experiment, and the average fidelity of single photon state after the resetting process for circuit I (II) is $F=0.870\pm0.012$ ($F=0.869\pm0.021$), and the entanglement fidelity for Circuit I (II) is $F=0.805\pm0.017$ ($F=0.807\pm0.033$). Although the experimental value are smaller than the theoretical prediction, in our opinion they are rational and reliable.

}

\section{Experimental results for $\mathcal{H}_0=\sigma_{y}/2$}

\begin{figure*}
	\centering
	\includegraphics[width=0.8\linewidth]{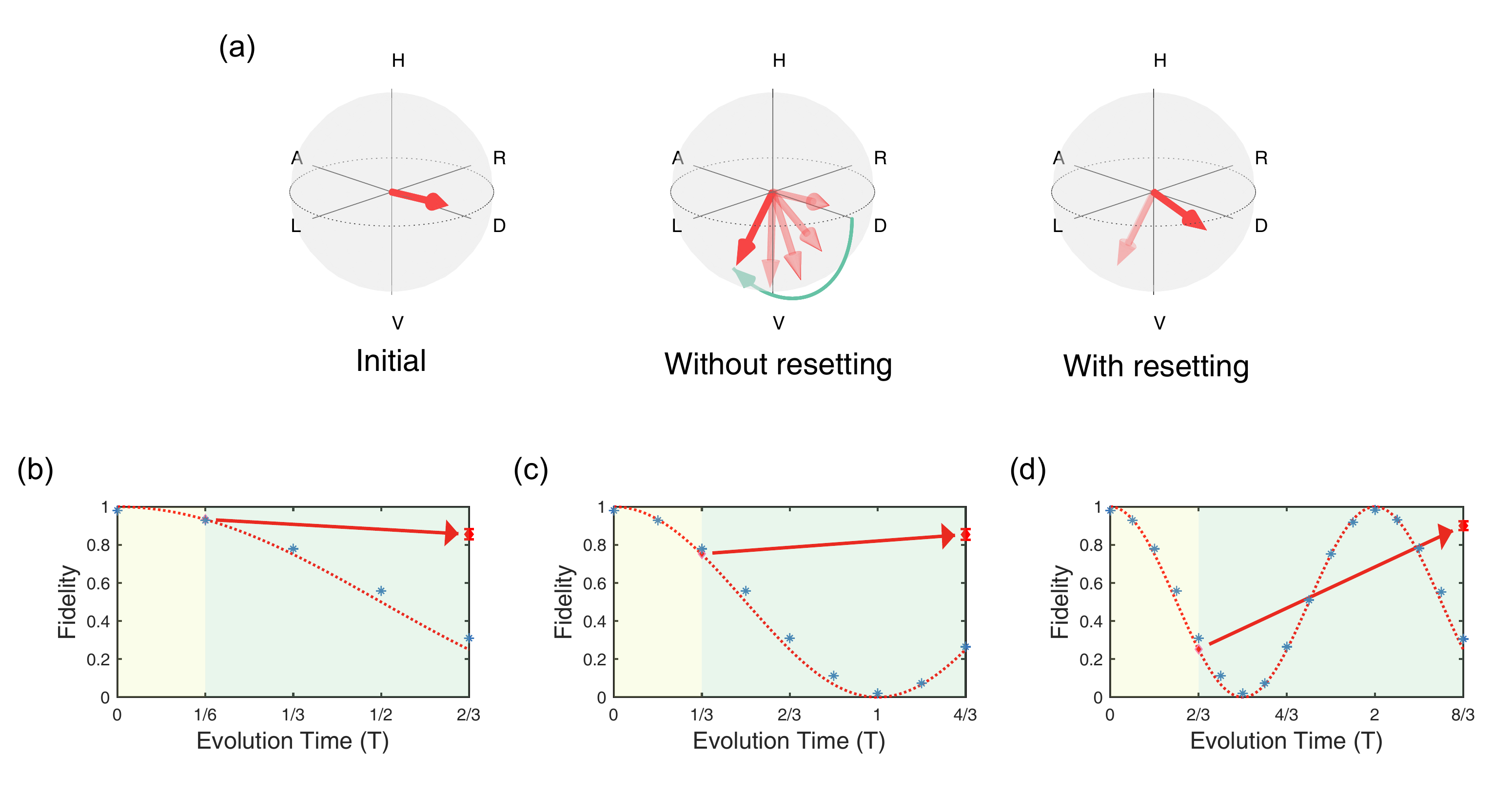}
	\caption{\textbf{Illustration of the quantum resetting process.} (a) Visual description of the evolution process before and after the resetting processing for circuit I with $\mathcal{H}_0=\sigma_{y}/2$. (b-d) The evolution of state fidelity during resetting process for circuit I with $\mathcal{H}_0=\sigma_{y}/2$.
	}
	\label{fig:fS1}
\end{figure*}

\begin{table*}[h!]
	\newcommand{\tabincell}[2]{\begin{tabular}{@{}#1@{}}#2\end{tabular}}
	\centering
	\begin{tabular}{|c|c|c|c|c|c|}
		\rowcolor{gray!10}
		\hline
		\tabincell{c}{Free\\Hamiltonian} & Interaction &  \tabincell{c}{Average fidelity\\ with different\\ Evolution time} & \tabincell{c}{Average fidelity\\ with different\\ initial states} & \tabincell{c}{Entanglement\\ fidelity} & \tabincell{c}{CHSH\\ value}\\
		\hline
		$\sigma_{z}/2$ & SWAP & $0.858\pm0.013$ & $0.870\pm0.012$ & $0.805\pm0.017$ & $2.13\pm0.05$\\
		\hline
		$\sigma_{y}/2$ & SWAP & $0.870\pm0.015$ & $0.874\pm0.016$ & $0.811\pm0.024$ & $2.19\pm0.05$\\
		\hline
		$\sigma_{z}/2$ & \tabincell{c}{$(I\otimes H)\cdot\text{G}_{\text{PBS}}$\\$\cdot(X\otimes I)$} & $0.858\pm0.024$ & $0.869\pm0.021$ & $0.807\pm0.033$ & $2.26\pm0.08$\\
		\hline
	\end{tabular}%
	\label{tab:summary}%
	\caption{\textbf{Summary of experimental results.}}
\end{table*}%

\begin{figure}
	\centering
	\includegraphics[width=0.9\linewidth]{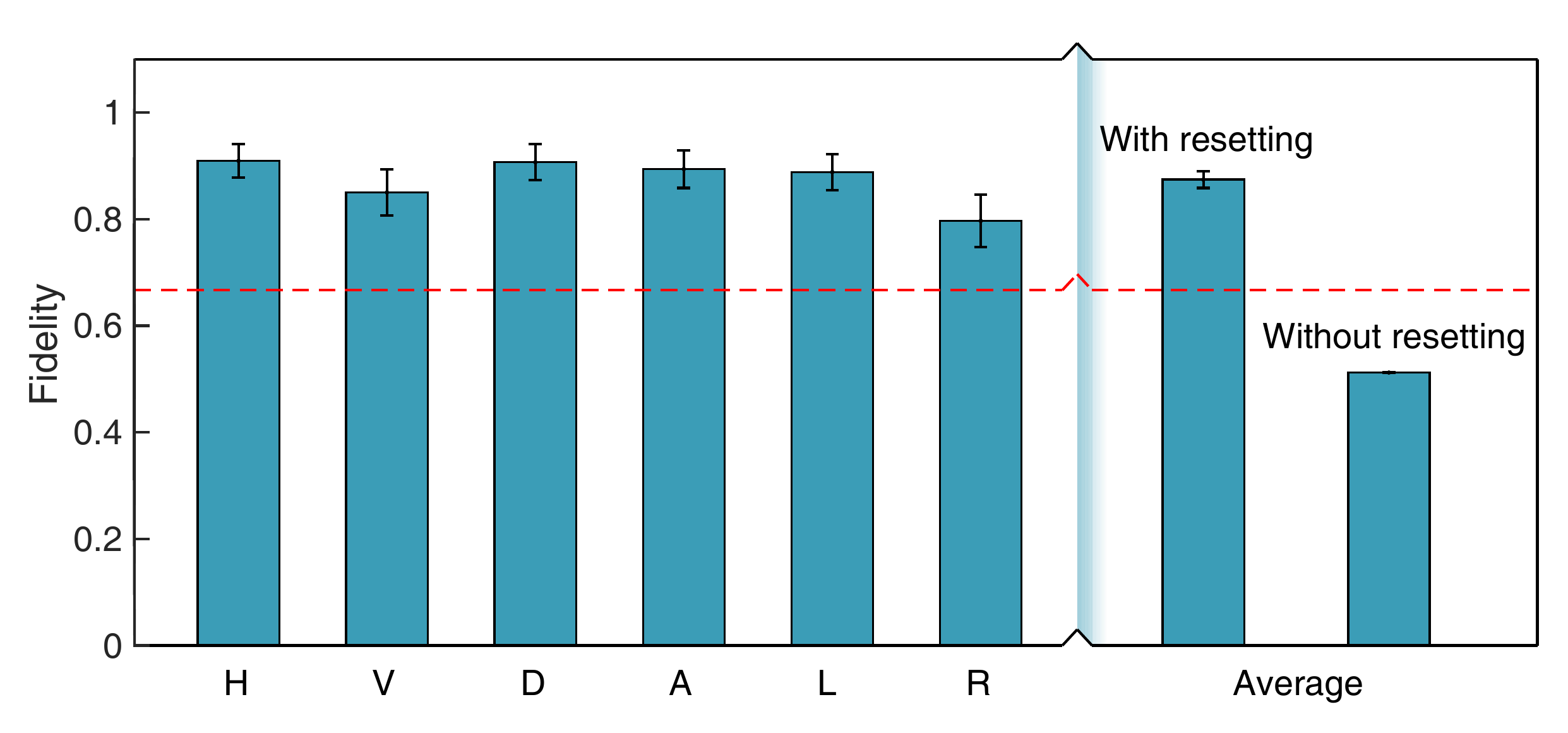}
	\caption{\textbf{The fidelity for different types of target states for circuit I with $\mathcal{H}_0=\sigma_{y}/2$.} With resetting, the average fidelity is $F=0.874\pm0.016$. Without resetting, the average fidelity for circuit I (II) is $F=0.5121\pm0.0003$.
	}
	\label{fig:fS2}
\end{figure}

\begin{figure}
	\centering
	\includegraphics[width=0.9\linewidth]{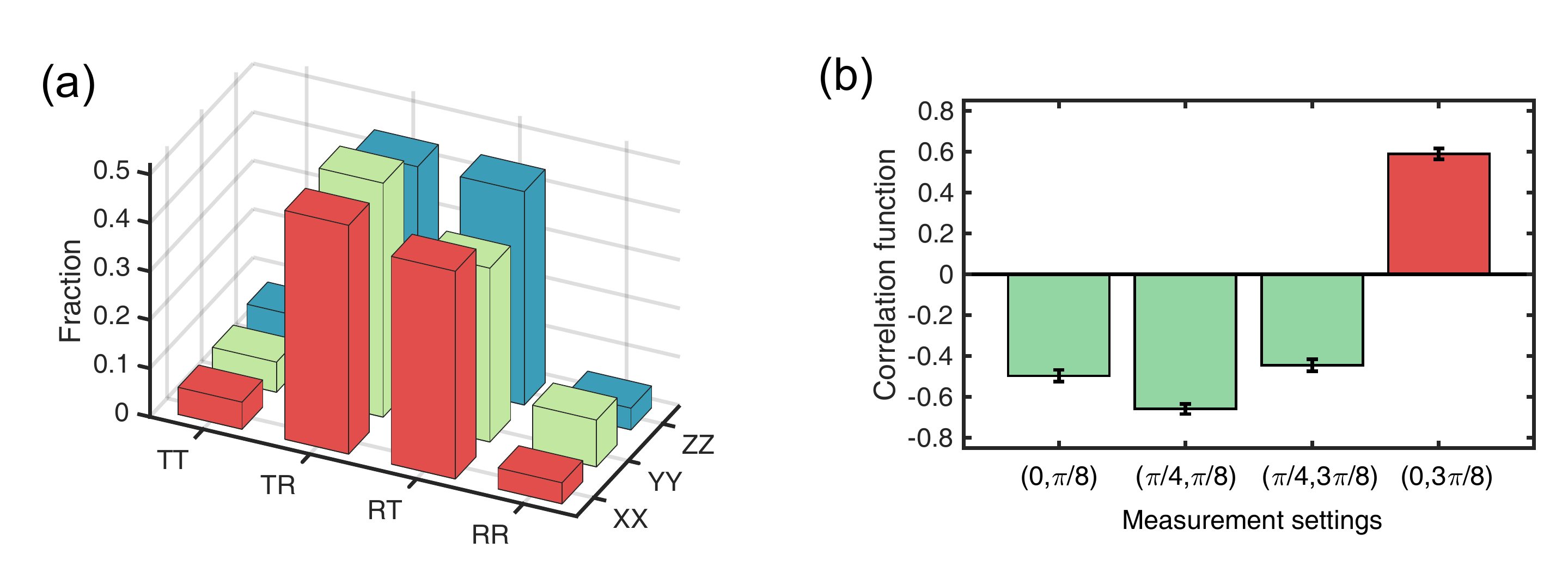}
	\caption{\textbf{The measured fraction and correlation of reset entanglement for circuit I with $\mathcal{H}_0=\sigma_{y}/2$.}  (a) The entanglement fidelity is $F=0.811\pm0.024$. (b) CHSH inequality value is $2.19\pm0.05$.
	}
	\label{fig:fS3}
\end{figure}

To verify that our photonic time-warp machine could reset the target system which evolves with arbitrary free Hamiltonian, we also change the $\mathcal{H}_0$ in circuit I as the form of $\mathcal{H}_0=\sigma_{y}/2$.
In experiment, we first illustrate the process of free evolution with and without resetting in Fig.~\ref{fig:fS1}. Further, we test the reliability of our photonic time-warp machine with the target quantum system initialized at different types of states. The result is shown in Fig~\ref{fig:fS2}. At last, we verify the capability to reset a target system which entangled with other systems. As shown in Fig.~\ref{fig:fS3}, the fidelity of entangled state is $F=0.811\pm0.024$ and the CHSH inequality value is $2.19\pm0.05$.
These results demonstrate the validity of our photonic time-warp machine for arbitrary free Hamiltonians.

\section{Summary of experimental results}

For a clear illustration, we give the summary of our experimental results in Table.~S1. In our experiment, the photonic time-warp machine successfully reset the target system with different free Hamiltonians, different interactions, different initial states and different Evolution time. The all results prove that our photonic time-warp machine could reset a really uncontrolled target system with acceptable fidelity.

\end{document}